\documentclass[preprint,amsmath,amssymb]{revtex4}

\usepackage{graphicx}
\usepackage{bm}
\usepackage{subfigure}

\begin{document}
\title{Experimental evidence for a surface distribution
of two-level systems in superconducting lithographed microwave
resonators}
\begin{abstract}

We present measurements of the temperature-dependent frequency shift
of five niobium superconducting coplanar waveguide microresonators
with center strip widths ranging from 3~$\mu$m to 50~$\mu$m, taken
at temperatures in the range 100-800~mK, far below the 9.2~K
transition temperature of niobium. These data agree well with the
two-level system (TLS) theory. Fits to this theory provide
information on the number of TLS that interact with each resonator
geometry. The geometrical scaling indicates a surface distribution
of TLS, and the data are consistent with a TLS surface layer
thickness of order a few nm, as might be expected for a native oxide
layer.

\end{abstract}
\author{Jiansong Gao}
\affiliation{Division of Physics, Mathematics, and Astronomy,
California Institute of Technology, Pasadena, CA 91125}

\author{Miguel Daal}%
\affiliation{ Physics Department, University of California at
Berkeley, Berkeley, CA 94720}

\author{Bernard Sadoulet}%
\affiliation{ Physics Department, University of California at
Berkeley, Berkeley, CA 94720}

\author{Benjamin A. Mazin}
\affiliation{Jet Propulsion Laboratory, California Institute of
Technology, Pasadena, CA 91109}

\author{Peter K. Day}
\affiliation{Jet Propulsion Laboratory, California Institute of
Technology, Pasadena, CA 91109}

\author{Henry G. Leduc}
\affiliation{Jet Propulsion Laboratory, California Institute of
Technology, Pasadena, CA 91109}

\author {Anastasios Vayonakis}
\affiliation{Division of Physics, Mathematics, and Astronomy,
California Institute of Technology, Pasadena, CA 91125}

\author{Shewtank Kumar}
\affiliation{Division of Physics, Mathematics, and Astronomy,
California Institute of Technology, Pasadena, CA 91125}


\author{Jonas Zmuidzinas}
\affiliation{Division of Physics, Mathematics, and Astronomy,
California Institute of Technology, Pasadena, CA 91125}

\date{\today}

\maketitle

Superconducting microresonators have attracted substantial interest
for low temperature detector applications due to the possibility of
large-scale microwave frequency
multiplexing\cite{Mazin02,Day03,Schmidt03,Mazinthesis,Hahn06,
Mates08,Vardulakis08,Baselmans08,Hahn08}. Such resonators are also
being used in quantum computing experiments
\cite{Wallraff04,Martinis05,Silanpaa07} and for sensing
nanomechanical motion\cite{Regal08}. We previously reported that
excess frequency noise is universally observed in these resonators
regardless of the type of superconductor or substrate being used,
and suggested that two-level systems (TLS) in dielectric materials
may be responsible for this noise\cite{Gao07}. TLS effects are also
observed in superconducting qubits\cite{Martinis05}. The TLS
hypothesis is strongly supported by the observed temperature
dependence of the noise, and also by the observation of
temperature--dependent resonance frequency shifts that agree closely
with TLS theory\cite{Kumar08}. To make further progress, it is
essential to constrain the location of the TLS -- to determine
whether they exist in the bulk substrate or in surface layers,
perhaps oxides on the exposed metal or substrate surfaces, or in the
interface layers between the metal films and the substrate. In this
paper, we provide direct experimental evidence for a surface
distribution of TLS.

The unusual physical properties of amorphous materials
at low temperatures are consistent with a collection
of TLS with a wide range of excitation energies and relaxation
rates\cite{Phillips72,Anderson72}.
The TLS have electric dipole moments that couple to the
electric field $\vec E$ of our resonators.
For microwave frequencies, and at temperatures $T$ between 100~mK and 1~K,
the resonant interaction dominates over relaxation,
which leads to a temperature-dependent variation of the dielectric
constant given by\cite{Phillips87}
\begin{equation}
\frac{\Delta\epsilon}{\epsilon} = -\frac{2\delta}{\pi}\left[\mathrm{
Re}\Psi\left(\frac{1}{2} + \frac{1}{2\pi i}\frac{\hbar
\omega}{kT}\right)-\log\frac{\hbar\omega}{kT}\right]\label{eqn:epsT}
\end{equation}
where $\omega$ is the frequency, $\Psi$ is the complex digamma
function, and $\delta = \pi Pd^2/3\epsilon$
represents the TLS-induced dielectric loss tangent
at $T=0$ for weak non-saturating fields.
Here $P$ and $d$ are the two-level density of states and
dipole moment, as introduced by Phillips\cite{Phillips87}.

Eq.~(\ref{eqn:epsT}) has been used extensively to derive values of
$Pd^2$ in amorphous materials. If TLS are present in superconducting
microresonators, their contribution to the dielectric constant
described by Eq.~(\ref{eqn:epsT}) could be observable as a
temperature--dependent shift in the resonance frequency. Indeed, it
has recently been suggested that the small anomalous
low--temperature frequency shifts often observed in superconducting
microresonators may be due to TLS effects\cite{Kumar06,Barends07},
and in fact excellent fits to the TLS theory can be
obtained\cite{Kumar08}. Assuming that the TLS are uniformly
distributed in a volume $V_h$ of host material (e.g. a metal oxide
or the bulk substrate) which has a dielectric constant of
$\epsilon_h$, it can be shown that the fractional resonance
frequency shift is given by
\begin{equation}
\frac{\Delta f_r}{f_r} = - \frac{F}{2 }\, \frac{\Delta
\epsilon}{\epsilon} \label{eqn:f0T}
\end{equation}
where the filling factor $F$ is given by
\begin{equation}
F = \frac{\int_{V_h}\epsilon_h \vec{E}(\vec{r})^2
d\vec{r}}{\int_{V}\epsilon  \vec{E}(\vec{r})^2
d\vec{r}}=\frac{w^e_h}{w^e}. \label{eqn:F}
\end{equation}
The factor $F$ accounts for the fact that the TLS host material
volume $V_h$ may only partially fill the resonator volume $V$,
giving a reduced effect on the variation of resonance frequency.
According to Eq.~(\ref{eqn:F}), $F$ is the ratio of the electric
energy $w^e_h$ stored in the TLS-loaded volume to the total electric
energy $w^e$ stored in the entire resonator.

The key idea of the experiment described in this paper is to measure
$\Delta f_r/f_r$ of coplanar waveguide (CPW) resonators with
different geometries in order to obtain values of $F\delta$ for each
geometry. The frequency-multiplexed resonators are all fabricated
simultaneously and are integrated onto a single chip, and are
measured in a single cooldown. We can therefore safely assume that a
single value of the loss tangent $\delta$ applies for all resonator
geometries. This allows the variation of the filling factor $F$ with
geometry to be determined, providing information on the geometrical
distribution of the TLS. If TLS are in the bulk substrate with
dielectric constant $\epsilon_r$, Eq.~(\ref{eqn:F}) applied
to the CPW field distribution would yield a filling
factor $F\approx\epsilon_r/(\epsilon_r+1)$ that is
independent of the resonator's center strip width $s_r$. If
instead the TLS are in a surface layer, $F$ should be dependent on
the CPW geometry, scaling roughly as $1/s_r$.


We used a device with a 120~nm-thick Nb film deposited on a
crystalline sapphire substrate, patterned into five CPW
quarter-wavelength resonators with different geometries. Because Nb
has a critical temperature $T_c = 9.2$~K, the effect of
superconductivity on the
temperature dependence of the resonance frequency
is negligible for $T < 1$~K.
As shown in Fig.~1, each resonator is
capacitively coupled to a common feedline using a CPW coupler of
length $l_c\cong200~\mu$m and with a common center-strip width of
$s_c=3~\mu$m. The coupler is then widened into the resonator body,
with a center-strip width of $s_r=$ 3~$\mu$m, 5~$\mu$m, 10~$\mu$m,
20~$\mu$m or 50~$\mu$m, and a length of $l_r\sim5~$mm. The ratio
between center strip width $s$ and the gap $g$ in both the coupler
and the resonator body is fixed to 3:2, to maintain a constant
impedance of $Z_0 \approx 50~\Omega$. The resonance frequencies are
$f_r\sim6$~GHz, and the coupler is designed to have a coupling
quality factor $Q_c\sim 50,000$. The device is cooled in a dilution
refrigerator, and its microwave output is amplified using a
cryogenic high electron mobility transistor (HEMT) amplifier on the
4~K stage. The complex transmission $S_{21}$ through the device and
HEMT is measured using a vector network analyzer locked to a
rubidium frequency standard, and resonance frequencies are obtained
by fitting these data\cite{Mazinthesis,Kumar08}.


Fig.~2 shows the measured frequency shifts
$\Delta f_r/f_r$ for the five resonators as a function of temperature
over the range 100~mK to 800~mK.
Although all of the resonators display a common shape for the
variation of frequency with temperature, the magnitude of the effect varies
strongly with geometry.
As shown by the dashed lines in Fig.~2, fits to the
TLS model (Eq.~\ref{eqn:f0T}) generally agree quite well
with the data. The nonmonotonic variation of the
dielectric constant with temperature is familiar from the TLS
literature: $f_r$ increases ($\epsilon$ decreases) when $T>\hbar
\omega/2k$; a minimum in $f_r$ (a maximum in $\epsilon$) occurs
around $T=\hbar \omega/2k$; at lower temperatures ($T<100$~mK),
we would expect to see a decrease in $f_r$ (increase in $\epsilon$)
as indicated by the extrapolation of the fit.
The largest deviations from the TLS model
(about 4\%)  occur at the lowest temperatures,
and are likely due to TLS saturation effects. Indeed,
power-dependent frequency shifts of this size are expected
theoretically and have also been previously observed
experimentally\cite{Kumar08}. In this paper, we will ignore these
small effects and focus on the geometrical dependence.

With the exception of the $3~\mu$m resonator, the measured values of
$F\delta$ from the fits have to be corrected for the coupler because
the coupler's center strip width $s_c=3~\mu$m differs from that of
the resonator, $s_c\neq s_r$. In the limit $l_c << l_r$, it can be
shown that the corrected filling factor is given by $F^*  = ( F -
tF_{3\mu \textrm{m}})/(1-t)$,
where $t =2 l_c/(l_c+l_r)$. The values of $F^*\delta$ are listed in
Table~1, as well as the ratios relative to the value for 3~$\mu$m
resonator.


A portion of the CPW inductance per unit length is contributed by
the kinetic inductance of the superconductor. The kinetic inductance
fraction, $\alpha$, depends on CPW geometry and may be determined by
measuring resonance frequency shifts at higher temperatures, closer
to $T_c$\cite{Gao06b}. We therefore measured the resonance
frequencies at 4.2~K ($0.46~T_c$), allowing the shift $\Delta f_r
(4.2~K) = f_r(4.2~{\rm K}) - f_r(100~{\rm mK})$ as well as the
kinetic inductance fraction to be calculated for each geometry, as
shown in Table~1.

Fig.~3 shows the results for the geometrical scaling of
the corrected filling factor $F^*$ and the kinetic inductance
fraction $\alpha$, plotted as ratios relative to their respective
values for the resonator with a $3~\mu$m wide center strip.
The observed strong variation of $F^*$ with geometry immediately
rules out a volume TLS distribution, and favors a surface distribution.
We investigate this in more detail by comparing the data
to two theoretically calculated  geometrical factors $g_{\mathrm{m}}$
and $g_{\mathrm{g}}$, which have units of
inverse length and are calculated from contour integrals in a
cross--sectional plane given by $g_{\mathrm{m}} =
\int_{\mathrm{metal}} \vec{E}^2 dl / V^2$ and $g_{\mathrm{g}} =
\int_{\mathrm{gap}} \vec{E}^2 dl/V^2$,
where $V$ is the CPW voltage. The first integral is actually a sum
of three contour integrals, taken over the surfaces of the three
metal conductors, the center strip and the two ground planes. The
second contour integral is taken over the two ``gaps'', the
surface of the exposed substrate in between the conductors. These
contours are illustrated in the inset of Fig.~1.
The integrals are evaluated numerically using the electric field derived
from a numerical conformal mapping solution to the Laplace equation.

According to Eq.~(\ref{eqn:F}),
$F^*$ should have the same scaling as $g_{\mathrm{m}}$ if the TLS
are distributed on the metal surface (or at the metal-substrate
interface), or as $g_{\mathrm{g}}$ if the TLS are located on the
surface of the exposed substrate. The kinetic
inductance of the CPW may also be calculated using a contour integral
similar to that of $g_{\mathrm{m}}$, except that the integrand is
replaced by $\vec{H}^2$\cite{Gao06b}. Because the magnetic field
$\vec{H}$ is proportional to $\vec E$ for a quasi-TEM mode, we
expect the kinetic inductance fraction $\alpha$ to have the same
geometrical scaling as $g_{\mathrm{m}}$.

Fig.~3 shows that the four quantities, $F^*$, $\alpha$,
$g_{\rm{m}}$ and $g_{\mathrm{g}}$, all scale as $s_r^{-\gamma}$ with
$\gamma=0.85 - 0.91$. The finite thickness of the superconducting
film is responsible for the deviations from $\gamma=1$. This is very
strong evidence that the TLS have a surface distribution and are not
uniformly distributed in the bulk substrate. Our data cannot
discriminate between a TLS distribution on the metal surface from a
TLS distribution on the exposed substrate surface (the gap), because
the corresponding theoretical predictions ($g_{\mathrm{m}}$ and
$g_{\mathrm{g}}$) are very similar and both agree with the data.
Future measurements of resonators with various center strip to gap
ratios may allow these two TLS distributions to be separated.


The absolute values of
$F^*\delta$ are also of interest. Assuming a typical value of $\delta \sim
10^{-2}$ for the TLS--loaded material\cite{Martinis05}, the measured
value of $F^*\delta=3\times10^{-5}$ for the 3~$\mu$m resonator
yields a filling factor of $F^* \sim 0.3\%$. Numerical calculations
show that this is consistent with
a $\sim2~$nm layer of the TLS--loaded material on the metal surface
or a $\sim3~$nm layer on the gap surface, suggesting that native
oxides or adsorbed layers may be the TLS host material.

In summary, the anomalous low--temperature frequency shifts of our
superconducting CPW resonators are well explained by a model in
which TLS are distributed on the surface of the CPW. The excess
frequency noise\cite{Day03,Gao07,Kumar08} also displays a strong
geometrical dependence: for a fixed internal power, we find that the
noise scales as $1/s_r^{1.6}$, consistent with a surface
distribution of TLS fluctuators\cite{Gao08}. It therefore seems very
likely that the TLS causing the anomalous low--temperature frequency
shifts are also responsible for the excess frequency noise. The use
of optimized geometries or non--oxidizing materials may therefore
offer a route to more sensitive photon detectors. The TLS should
also affect the resonator dissipation. At the relatively high power
levels used in our experiments, TLS dissipation is strongly
saturated\cite{Martinis05,Gao07} and rather high values of the
resonator quality factor ($Q_r>10^5$) are routinely
obtained\cite{Mazin02}. However, at low enough microwave power, the
TLS response should become unsaturated\cite{Martinis08} at which
point the quality factor should be  limited to $Q_r \sim 1/F
\delta$, or around $3 \times 10^4$ for the $3~\mu$m resonator
described in this paper. This low--power regime is of direct
relevance for quantum experiments\cite{Wallraff04}, in which the
microwave excitation of the resonator consists of one or a few
photons.

We thank John Martinis, Clare Yu and Sunil Golwala for useful
discussions. The device was fabricated in the University of
California, Berkeley, Microfabrication Laboratory. This work was
supported in part by the NASA Science Mission Directorate, JPL, the
Gordon and Betty Moore Foundation, and Alex Lidow, a Caltech
Trustee.

\newpage

\newpage
\begin{table}[h]
\caption{Values and ratios}
\begin{tabular}{ccccccccc}
\hline
$s_r$&$\Delta f_r(4.2~\mathrm{K})$&$\frac{\alpha}{\alpha_{3\mu\mathrm{m}}}$ & $F^*\delta$~&~$\frac{F^*}{F^*_{3\mu\mathrm{m}}}$&$\frac{g_{\mathrm{m}}}{g_{\mathrm{m},3\mu\mathrm{m}}}$&$\frac{g_{\mathrm{g}}}{g_{\mathrm{g},3\mu\mathrm{m}}}$\\
$[\mu$m]&[MHz]&&$\times10^{-5}$&&&\\
\hline
3~$\mu$m    &  11.1   &1        &$2.98\pm0.12$  &1      &1      &1\\
5~$\mu$m    &  7.41   &0.67     &$2.00\pm0.07$  &0.67   &0.62   &0.64\\
10~$\mu$m   &  4.15   &0.37     &$1.10\pm0.03$  &0.37   &0.33   &0.35\\
20~$\mu$m   &  2.28   &0.21     &$0.54\pm0.03$  &0.18   &0.17   &0.19\\
50~$\mu$m   &  1.02   &0.092    &$0.24\pm0.02$  &0.08   &0.075  &0.086\\
\hline



\end{tabular}\\
\label{tab:ratios}
\end{table}

\newpage
\begin{center}
FIGURE CAPTIONS
\end{center}
\vspace{1cm}

\noindent
Figure 1\\
(Color online) An illustration of the CPW coupler and
resonator. The inset shows a cross-sectional view of the CPW. The contour
of the metal surface and the contour of the exposed surface of
the substrate are indicated by the solid line and the dashed line,
respectively.
\vspace{0.5cm}

\noindent
Figure 2\\
(Color online) Fractional frequency shift $\Delta f_r/f_r$ as a
function of temperature. $\Delta f_r/f_r$ is calculated using
$\Delta f_r/f_r = [f_r(T) -
f_r(800~\mathrm{mK})]/f_r(800~\mathrm{mK})$. The temperature sweep
is in steps of $50~$mK from 100~mK to 600~mK, and in steps of 100~mK
above 600~mK. The markers represent different resonator geometries,
as indicated by the values of the center strip width $s_r$ in the
legend. The dashed lines indicate fits to the TLS theory
(Eq.~\ref{eqn:f0T}). \vspace{0.5cm}

\noindent
Figure 3\\
(Color online) The scaling of the measured values of the
kinetic inductance $\alpha$ and TLS filling factor $F^*$, as well as
the calculated values of the CPW geometrical factors $g_{\rm{m}}$
and $g_{\rm{g}}$, are shown, as a function of the resonator center
strip width $s_r$. The top panel shows the ratios of $\alpha$
($r_1$), $F^*$ ($r_2$), $g_{\rm{m}}$ ($r_3$) and $g_{\rm{g}}$
($r_4$) to their values for the 3~$\mu$m resonators. The bottom
panel shows these ratios normalized by the kinetic inductance ratio
$r_1$.

\newcommand{\placefigures}{
\newpage
\begin{figure}
  \begin{center}
  \resizebox{6.5in}{!}{\includegraphics{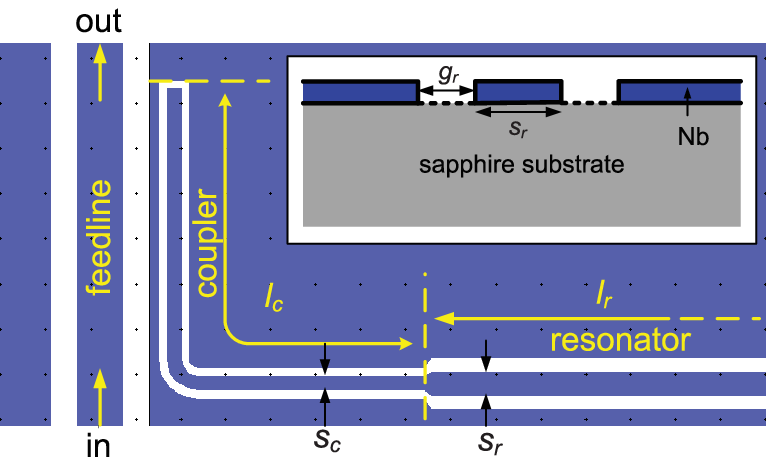}}\\
\vspace{1in} Figure 1
  \end{center}
\end{figure}
\newpage
\begin{figure}
  \begin{center}
  \resizebox{6.5in}{!}{\includegraphics{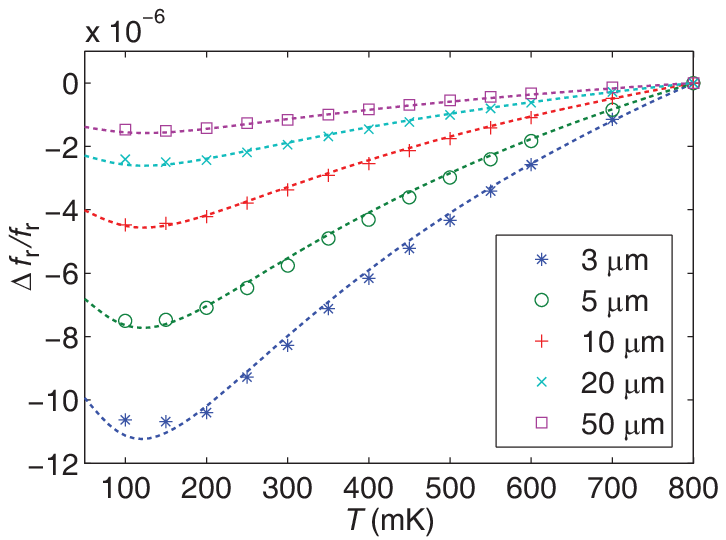}}\\
\vspace{1in} Figure 2
  \end{center}
\end{figure}
\newpage
\begin{figure}
  \begin{center}
  \resizebox{6.5in}{!}{\includegraphics{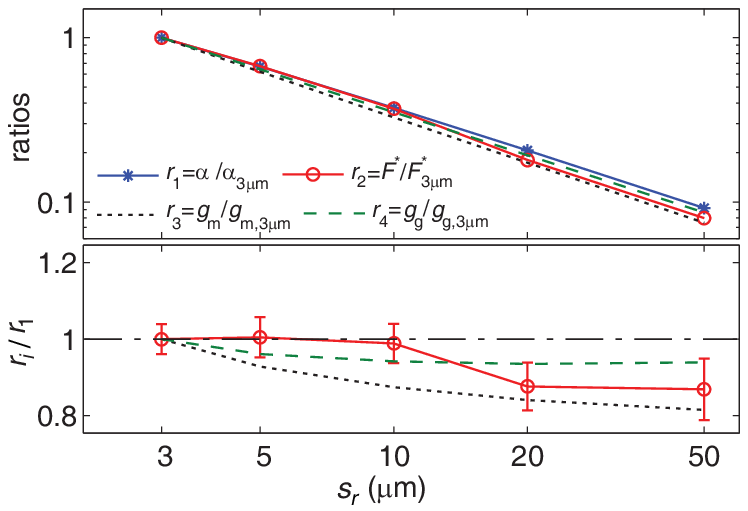}}\\
\vspace{1in} Figure 3
  \end{center}
\end{figure}
}
\placefigures

\end{document}